\documentclass[aps,prl,showpacs,twocolumn]{revtex4-2}
\usepackage{graphicx}
\usepackage{graphicx}
\usepackage{amsmath}
\usepackage{amsfonts}
\usepackage{xcolor}

\begin{document}
	
	\title{Spectral Network Principle for Frequency Synchronization in Repulsive Laser Networks}

	\author{Mostafa Honari-Latifpour,$^{1,2}$ Jiajie Ding,$^{1,2}$ Igor Belykh,$^{3}$ and Mohammad-Ali Miri$^{1,2,*}$}
	
	\address{$^1$Department of Physics, Queens College, City University of New York, New York, New York 11367, USA\\
		$^2$Physics Program, The Graduate Center, City University of New York, New York, New York 10016, USA\\
		$^3$Department of Mathematics \& Statistics and Neuroscience Institute, Georgia State University, P.O. Box 4110, Atlanta, Georgia, 30302-410, USA}	
	\begin{abstract}
		Network synchronization of lasers is critical for reaching high-power levels and for effective optical computing. Yet, the role of network topology for the frequency synchronization of lasers is not well understood. Here, we report our significant progress toward solving this critical problem for networks of heterogeneous laser model oscillators with repulsive coupling. We discover a general approximate principle for predicting the onset of frequency synchronization from the spectral knowledge of a complex matrix representing a combination of the signless Laplacian induced by repulsive coupling and a matrix associated with intrinsic frequency detuning.  We show that the gap between the two smallest eigenvalues of the complex matrix generally controls the coupling threshold for frequency synchronization. In stark contrast with Laplacian networks, we demonstrate that local rings and all-to-all networks prevent frequency synchronization, whereas full bipartite networks have optimal synchronization properties. Beyond laser models, we 
		show that, with a few exceptions, the spectral principle can be applied to repulsive Kuramoto networks. Our results may provide guidelines for optimal designs of scalable laser networks capable of achieving reliable synchronization.
	\end{abstract}
	%\pacs {05.45.-a}
	
	\date{\today}
	\draft \maketitle
	
	{\it Introduction.} 
	Frequency synchronization when coupled photonic oscillators with different natural frequencies synchronize to a common frequency 
	is a critical requirement for unconventional computing using lasers \cite{nixon2013observing,eckhouse2008loss,honari_2020,parto2020realizing,Fienup:82} or trapped Bose-Einstein condensates \cite{berloff2017realizing} as well as for high-power beam combining for communication, sensing, and metrology \cite{leger1994diode}. Complex laser oscillator networks that expand beyond the conventional lattice geometries based on the evanescent tail coupling of the neighboring lasers can be implemented using diffraction engineering \cite{PhysRevLett.108.214101,PhysRevLett.106.223901,Brunner:15}. The main types of coupling in laser networks encompass dispersive and dissipative interactions.
	Dissipative coupling induces the splitting of the resonant frequencies and is generally considered the superior mechanism for promoting network synchronization \cite{ding2019dispersive}. However, the dissipative coupling can be attractive or repulsive, promoting in-phase and out-of-phase oscillations, respectively. The significance of the repulsive coupling scenario manifests itself in various applications, including the spin models for unconventional computing \cite{nixon2013observing,10.1007/978-3-030-19311-9_19,wang2013coherent}. In this context, the 
	attractive coupling corresponds to the trivial ferromagnetic case, whereas repulsive coupling aligns with anti-ferromagnetism that can embed hard optimization problems \cite{berloff2017realizing,honari_2020, 10.3389/fphy.2014.00005}, and can represent non-trivial energy based neural network models \cite{MiriMenon+2023+883+892}. Furthermore, it has been suggested that anti-phase-coupled lasers can  have better overall beam combining efficiencies \cite{Andrianov:20}.
	
	Extensive research has been devoted to the role of network structure and parameter heterogeneity on the synchronization in oscillator networks with attractive coupling, including laser arrays \cite{braiman1995taming, PhysRevA.47.4287,PhysRevE.55.3865,kozyreff2000global,zamora2010crowd,mahler2020experimental,nair2021using}, and more broadly, Laplacian \cite{pecora1998master,boccaletti2002synchronization,belykh2004connection,sun2009master,nishikawa2010network,boccaletti2006complex}, pulse-coupled \cite{belykh2005synchronizationOfBursting,belykh2015synergistic}, and Kuramoto-type networks \cite{acebron,restrepo2005onset,arenas2006synchronization,dorfler2014synchronization,medvedev2015stability,rodrigues2016kuramoto,nishikawa2016symmetric,zhang2021random,skardal2014optimal,zhang2017identical}. Yet, a significant knowledge gap remains regarding the interplay of these factors for frequency synchronization in repulsive oscillator networks. Such networks exhibit different forms of frequency synchronization, including splay states \cite{berner2021generalized}, clusters \cite{ronge2021splay}, and cyclops states \cite{munyayev2023cyclops} whose dependence on the network structure is not well understood  and can be counterintuitive. For example, globally coupled repulsive Kuramoto networks fail to reach frequency synchronization whereas it occurs in locally coupled networks \cite{tsimring2005repulsive}.
	
	In this Letter, we discover a general principle that pairs frequency synchronization with the network structure and parameter detuning in networks of class-A laser oscillators \cite{ARECCHI1984308,PhysRevA.43.4997} with repulsive signless Laplacian dissipative coupling \cite{ding2019dispersive}. Much in the vein of the master stability function for complete synchronization in Laplacian networks \cite{pecora1998master,sun2009master}, this principle can predict a coupling threshold for frequency synchronization from the spectral knowledge of the complex matrix composed of the connectivity matrix and the matrix representing intrinsic frequency detuning. In contrast to complete synchronization in Laplacian networks, the coupling threshold in such laser networks is generally controlled by the spectral gap between the two smallest (non-zero) eigenvalues of the complex matrix. This principle suggests that full bipartite networks rather than global or local network topologies provide optimal synchronization properties.\\
	\begin{figure}
		\label{fig1}
		\centering
		\includegraphics[width=0.5\textwidth]{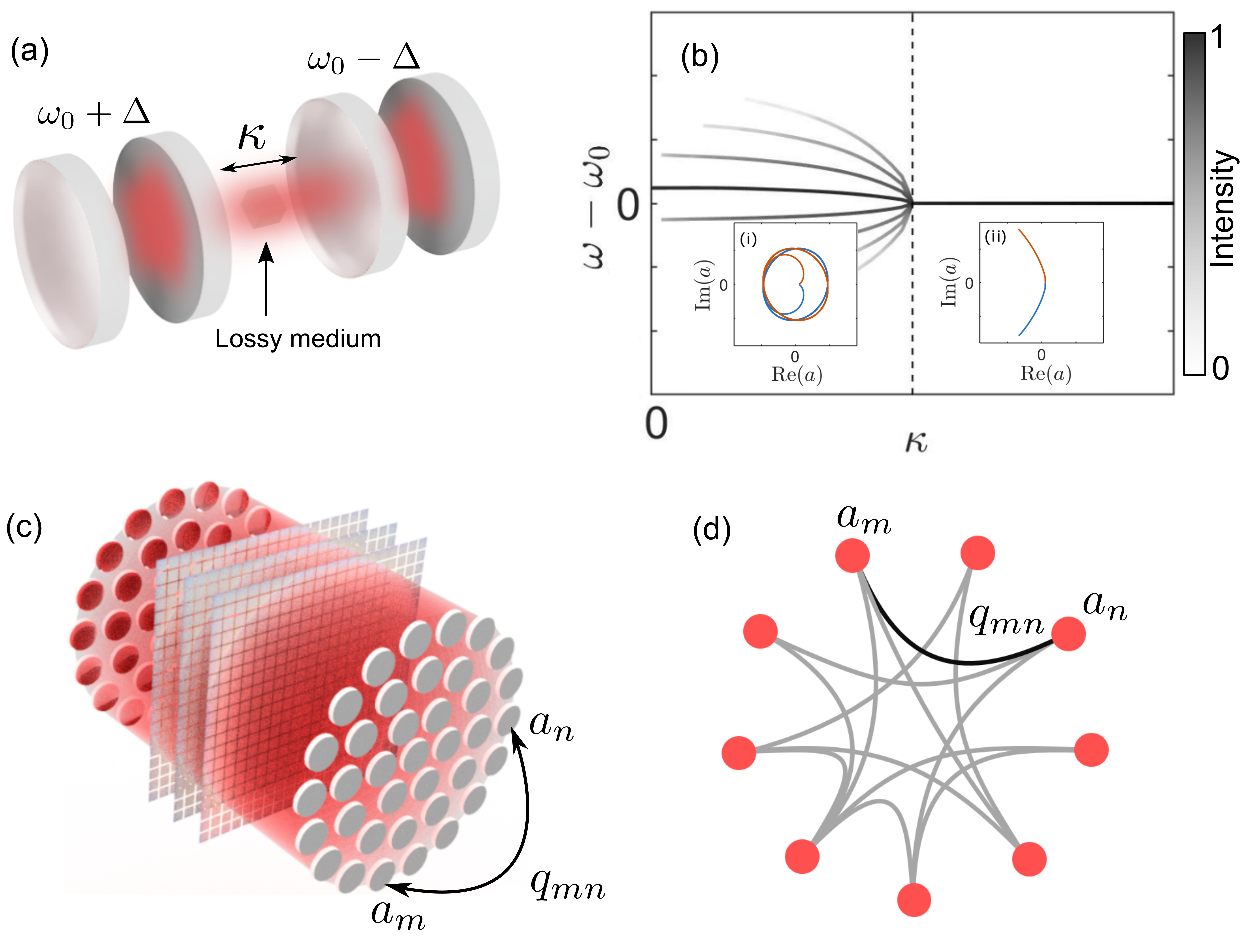}
		\caption{(a) The concept of dissipative coupling, where the interaction between two laser cavities is mediated through a dissipative medium which can promote anti-phase synchronization (repulsive coupling). (b) 	The frequency spectrum of the two detuned laser oscillators from (a) and the transition to the anti-phase frequency synchronization via increasing the coupling $\kappa$. The insets show the phase portraits $\Re(a_1),\Im(a_1)$ (red) and $\Re(a_2),\Im(a_2)$ (blue) before and after the critical phase transition. Parameters are $\omega_0=1,$ $\Delta=0.005,$ and $g_0=0.02.$ (c) General scheme for creating arbitrary coupling between laser oscillators by diffraction engineering \cite{PhysRevLett.108.214101}. (d) The equivalent network graph.}
	\end{figure}
	
\begin{figure}
		\includegraphics[width=0.9\columnwidth]{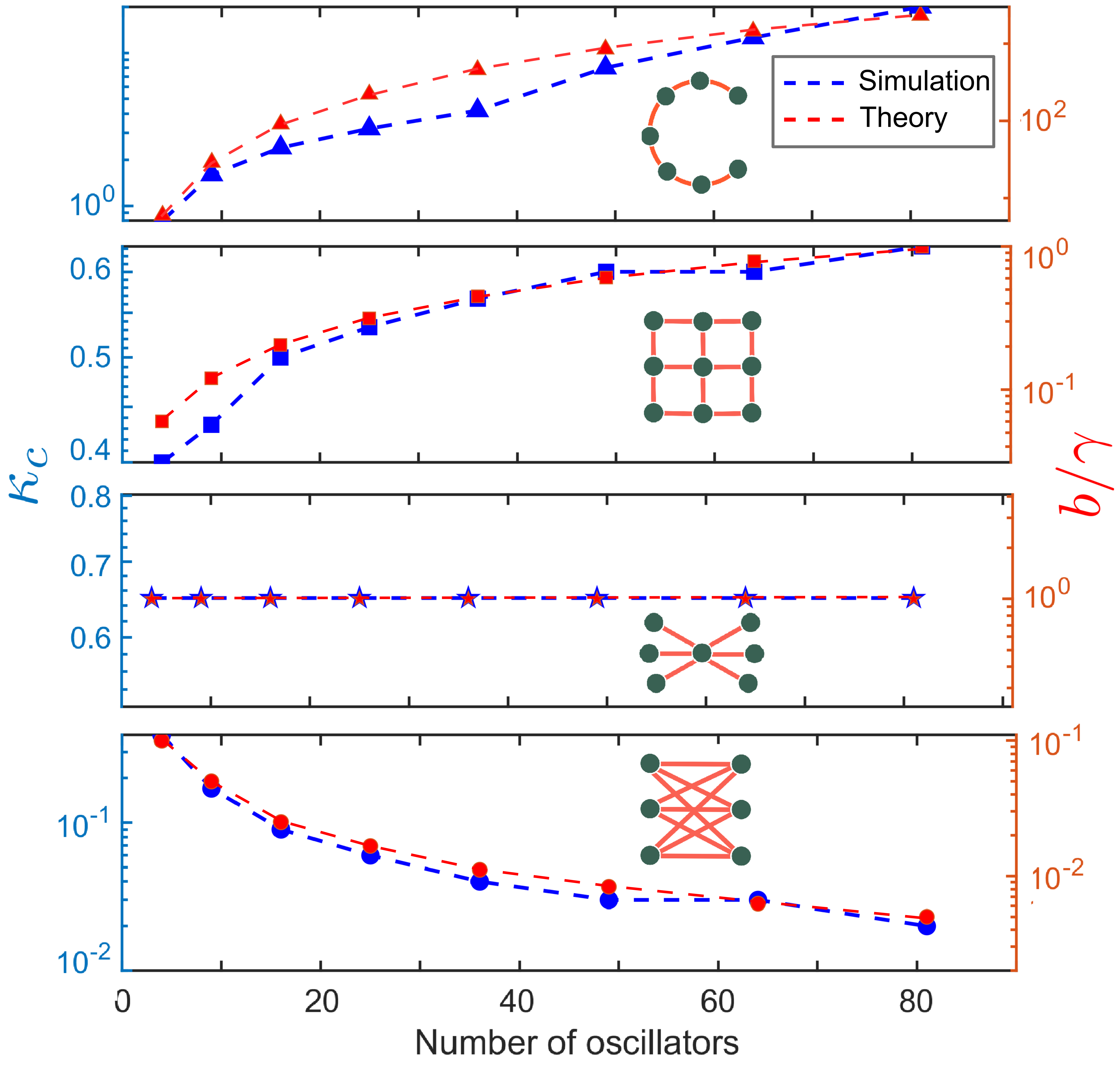}
		\caption{Actual (blue) and predicted (red) frequency synchronization threshold  $\kappa_c$ (with the order parameter $R>0.99$) in common bipartite network topologies, ranging from sparse to dense graphs. The predicted thresholds are computed from the spectral principle \eqref{k_c} with the scaling parameter $b$ chosen to fit the data. Parameters are $\omega_0=1$ and $g_0=0.02.$}\label{bipartite}
	\end{figure}

\begin{figure}
		\centering
		\includegraphics[width=0.9\columnwidth]{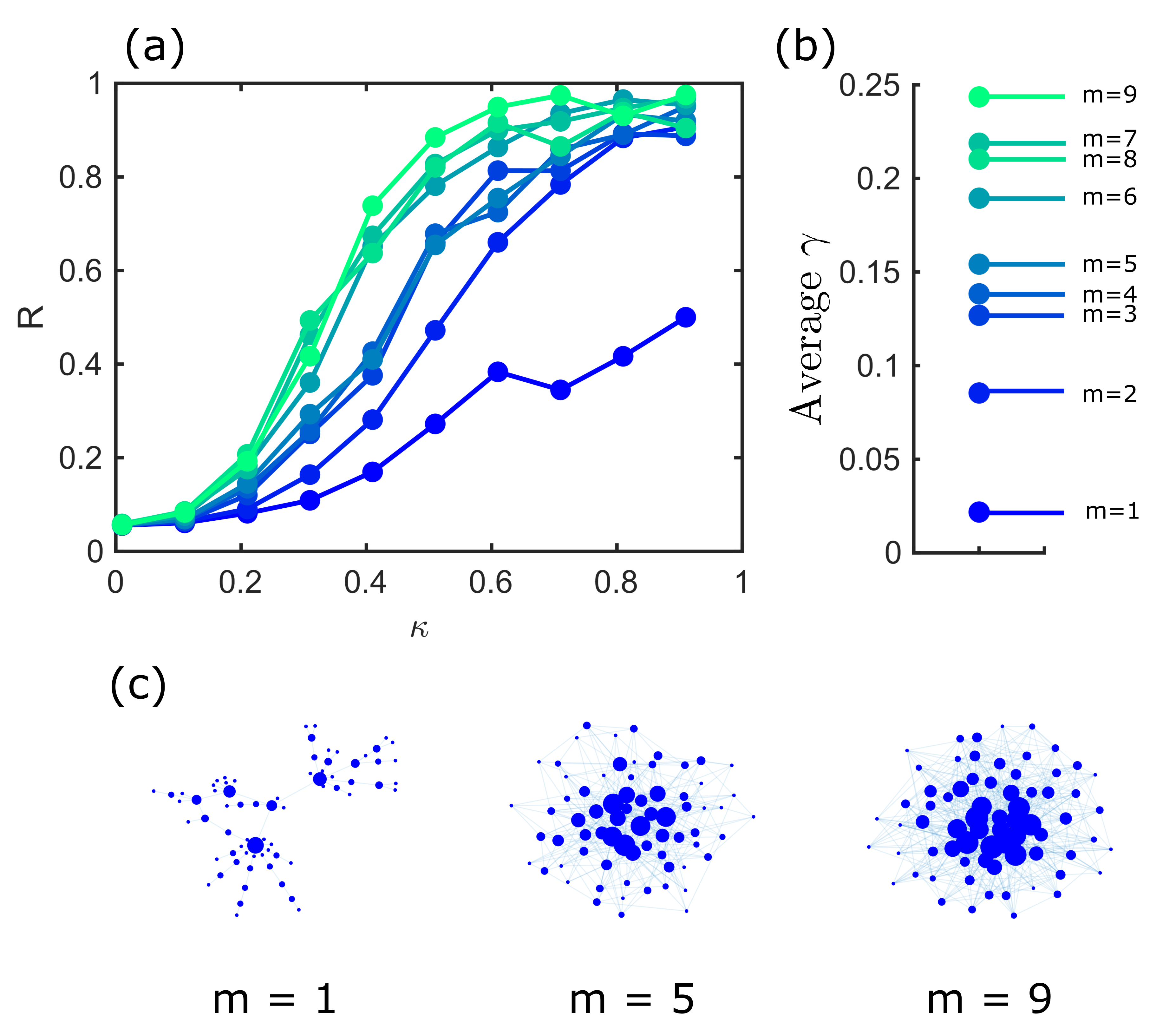}
		\caption{\label{fig6}Frequency synchronization of scale-free networks of identical oscillators and its relation to the spectral gap $\gamma$. The scale-free networks are generated from an initial graph with $m+10$ nodes via the preferential attachment mechanism
			\cite{RevModPhys.74.47}. (a). The onset of frequency synchronization via the dependence of the order parameter $R$ on coupling strength $\kappa.$ (b). The corresponding average spectral gap, $\gamma$, for networks with different $m$. Each curve in (a) and point in (b) correspond to the average of 100 randomly generated graphs of size $N=70$ with $m =1,...,9$. Other parameters are as in Fig.~\ref{bipartite}. A larger spectral gap enhances network synchronizability.  (c). Sample networks with $m=1,5,9.$}
	\end{figure}

	{\it Model formulation.} We consider a network of $N$ dissipatively coupled lasers described  by a minimal dynamical model that involves only the amplitude and phase of the field in each laser cavity \cite{born_wolf_1999}. The complex amplitude of the $n$th oscillator obeys $\dot{a}_n(t) = (-i\omega_n - 1 + g_0(1-|a_n|^2))a_n, $ $n=1,...,N,$ where time is normalized to the field decay rate, $\omega_n$ and $g_0$ represent the dimensionless resonant frequency and small signal gain, respectively. This model is valid when the atomic degrees of freedom are adiabatically eliminated for the so-called class-A lasers \cite{ARECCHI1984308,PhysRevA.43.4997}. Its individual dynamics is similar to that of the Landau-Stuart oscillator. The dynamical equations governing complex field amplitudes of the network are
	\begin{equation}\label{model}
		\dot{\mathbf{a}}(t) = -\mathbf{a} + g_0 ( 1 - {\mathbf{a}}^{*} \cdot \mathbf{a} ) \cdot \mathbf{a} -i \Omega \mathbf{a} - \kappa Q\mathbf{a},
	\end{equation}
	where $\mathbf{a}=[a_1,...,a_N]^T$ is the vector containing the lasers complex amplitudes, $\Omega=\mbox{\rm diag}(\omega_1,...,\omega_N)$ is an $N\times N$ diagonal matrix involving detuned resonant frequencies, $Q$ is the signless Laplacian connectivity matrix with off-diagonal elements $q_{mn}=1$ for coupled oscillators and $q_{mn}=0$ otherwise, and diagonal elements $q_{mm}=\sum_ {m\neq n} q_{mn}$. The negative sign of the coupling term $-\kappa Q\mathbf{a}$ with
	coupling coefficient $\kappa>0$ determines the repulsive nature of the dissipative coupling. Combining the last two terms in \eqref{model}, we 
	introduce the complex matrix %$ H = \Omega - i \kappa Q$,
	\begin{equation}
		M = i \Omega+ \kappa Q,
	\end{equation}
	which accounts for the contribution of intrinsic frequency detuning and linear coupling. In an amplitude and phase representation of the complex amplitudes $a_n(t)=A_n(t)e^{i \phi_n(t)}$, the network \eqref{model} can be written in the form 
	\begin{equation}
		\begin{array}{l}
			\dot{A}_n=-1+ g_0(1-A^2_n)A_n-\kappa \sum\limits_{m=1}^N  q_{mn} \cos(\phi_m-\phi_n),\\
			\dot{\phi}_n=-\omega_n-\kappa \sum\limits_{m=1}^N  q_{mn} \frac{A_n}{A_m}\sin(\phi_m-\phi_n),\;n=1,...,N.	
		\end{array}
		\label{real}
	\end{equation}
	Under the simplifying assumption that the amplitudes of all laser oscillators settle down to the same value so that $A_n(t) \rightarrow 1,$ the system \eqref{real} can be reduced to the classical repulsive Kuramoto model with an arbitrary adjacency matrix $C=Q-q_{mm}I,$ where $q_{mm}I$ is the degree matrix of the connection graph. However, the dynamics of the Kuramoto model and the full system \label{eq:2} can be different.\\
	{\it The spectral network principle.} Frequency synchronization occurs in the network \eqref{real} when $\langle \dot{\phi}_1 \rangle =	\langle \dot{\phi}_2 \rangle=...=\langle \dot{\phi}_N \rangle, $ where  $\langle ... \rangle$ denotes a time average. Hereafter, we will be using an order parameter $R =  \frac{2}{n(n-1)} \langle \sum_{i<j}^n \exp\{-(\dot\phi_i - \dot\phi_j)^2\} \rangle$ as a measure for the degree of frequency coherence, with $R=1$ corresponding to perfect frequency synchronization. Previous studies used energy Lyapunov-type functions to derive
	conditions on the stability of frequency synchronization in the classical Kuramoto model with global attractive coupling \cite{dorfler2014synchronization} and local repulsive coupling \cite{tsimring2005repulsive}. However, constructing such functions for the amplitude-phase model \eqref{model} with arbitrary repulsive coupling is elusive.  Here, we use an alternative approach to making sense of the complex matrix $ M$'s spectral properties as a network synchronizability criterion. We view the onset of frequency synchronization as competition between the network eigenmodes for oscillation. 
	This can be better understood in the case of identical oscillators, i.e., when $\omega_1 = \omega_2 = \cdots = \omega_N=\omega_0$. 
	In this case, considering the dynamics starting at low field intensities $ | \mathbf{a} | \ll \mathbf{1}$, the evolution can be linearized in the rotating frame of $\omega_0$ as 
	\begin{equation}
		\dot{\mathbf{a}} = (g_0 - 1) \mathbf{a} - \kappa Q\mathbf{a}.
		\label{identical}
	\end{equation}
	The connectivity matrix $Q$ has $N$ real eigenvalues $s_1\le s_2 \le ...\le s_N.$ Diagonalizing \eqref{identical} using the eigenmode basis of $Q$, $\mathbf{a}(t) = \sum_m \alpha_m(t) \mathbf{v}_m$, where $\alpha_m(t) = \mathbf{v}_m^{\dagger} \mathbf{a}(t),$
	we obtain the evolution equation for the $m$th eigenmode
	\begin{equation}
		\dot{\alpha}_m(t) = (g_0 - \kappa s_m - 1) \alpha_m,\;m=1,...,N.
		\label{diag}
	\end{equation}
	The fundamental mode ($m=1$) with the maximum net small-intensity gain, $g_0 - \kappa s_1 - 1,$ 
	has a higher probability of becoming the lasing mode, thereby inducing frequency synchronization. To do so, it needs to win the lasing competition with its closest competing mode with $m=2$ and the gain $g_0 - \kappa s_2 - 1.$ 
	The outcome of this competition is generally controlled by the gain difference between the two modes, $\kappa (s_2-s_1),$ that has to exceed an energy threshold. This suggests that the threshold coupling, $\kappa_c,$ for the onset of frequency synchronization can be estimated as 
	\begin{equation}
		\kappa_c =\frac{b}{s_2-s_1}\equiv \frac{b}{\gamma},
		\label{k_c}
	\end{equation}
	where $s_1$ and $s_2$ are the first and second smallest eigenvalues of the signless Laplacian matrix $Q,$ and the parameter $b$ is determined by the intrinsic properties of the individual laser oscillator and its lasing threshold. Note that the spectral gap $\gamma=s_2-s_1$ is zero for the globally coupled network \eqref{model}, so frequency synchronization cannot be achieved even for large values of $\kappa.$ This observation agrees with the similar property of the repulsive Kuramoto model of identical oscillators \cite{tsimring2005repulsive}. Similarly, networks with the zero spectral gap, $\gamma=0,$ are expected to be non-synchronizable. The property that guarantees a non-zero spectral gap is the bipartineness of the graph associated with the matrix $Q.$ 	It has been previously shown in the context of spectral signless Laplacian graph theory \cite{VANDAM2003241,cvetkovic2007signless,CVETKOVIC20102257} that the more edges one needs to remove to make the graph bipartite, the larger the smallest eigenvalue \cite{https://doi.org/10.1002/jgt.3190180210} and the smaller the spectral gap are. Therefore, a bipartite graph that generally is the easiest to synchronize has its smallest eigenvalue 
	at zero, leading to a larger spectral gap. To support this claim and validate the predictive power of the spectral network principle \eqref{k_c}, we numerically studied the scaling of the synchronization threshold $\kappa_c$ as a function of the network size in four common network topologies, ranging from sparse to dense graphs (Fig.~\ref{bipartite}).  All four types of networks discussed here are bipartite graphs and hence synchronizable. For all these networks, the spectral gap $\gamma$ can be calculated analytically as a function of $N.$
	For the chain graph $\gamma=s_2 - s_1 = 1 - \cos{(\pi/N)} $, which for large $N$ can be approximated as $\pi^2 / N^2$ \cite{Ding:19}. For the square lattice graph, the gap is $\gamma = 1 - \cos{(\pi/\sqrt{N})}$ and for large $N$ it is approximated by  $\pi^2 / n$. The star graph's gap is constant and equal to $1.$ Finally, for the full bipartite graph the gap is $\gamma= N/2$ for even $N$ and $\gamma= N-2$ for odd $N$. Figure~\ref{bipartite} indicates that the spectral network criterion \eqref{k_c} predicts the scaling of the coupling threshold rather precisely.
	To further illustrate the critical role of the spectral gap $\gamma$ in frequency synchronization, we  
	generated ensembles of uniformly connected, Barabasi-Albert scale-free networks \cite{RevModPhys.74.47}.  Figure \ref{fig6} shows that more heterogeneous networks with higher node degree hubs, in general, correspond to a larger spectral gap $\gamma$, and such networks are easier to synchronize.
	\begin{figure*}
	\centering
	\includegraphics[width=0.9\textwidth]{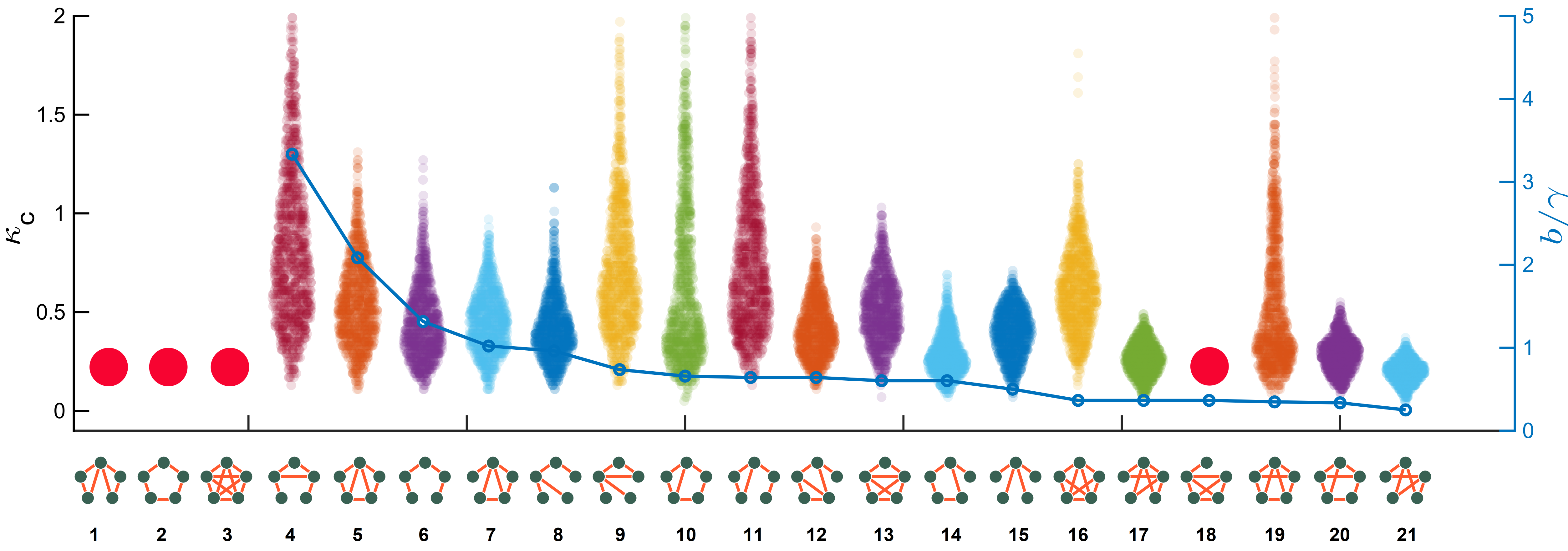}
	\caption{Synchronization threshold $\kappa_c$ for all $21$ possible connected networks of five detuned oscillators. 
		The scattered points in each violin plot represent the coupling thresholds for $1,000$ random frequency distributions with $\omega_m \in \mathcal{U}(-0.5,0.5).$ The corresponding network is shown under each plot. The large red circles indicate an infinite coupling threshold corresponding to non-synchronizable networks. The networks are ordered from $1$ to $21$ by the spectral gap $\gamma.$ The full bipartite network with index $21$ has the largest $\gamma.$ As a reference, the blue line shows a predicted trend from the spectral criterion for identical oscillators \eqref{k_c}, with the scaling constant $b$ calculated from the lowest value of $\kappa_c$ for full bipartite network $21.$}\label{fig4}
\end{figure*}\\
	{\it Extension to nonidentical laser oscillators.}
	While the criterion \eqref{k_c} performs remarkably well for a large spectrum of the regular and scale-free networks depicted in  Figs.~\ref{bipartite}-\ref{fig6}, it is important to point to the limitations of its predictive power. The energy landscape governing the system of repulsively coupled identical oscillators via a hypothetical Lyapunov function
	may be a non-convex function. As a result,  the fundamental eigenmode might not necessarily be the one to win the lasing competition or have the maximal overlap with the lasing mode. Therefore, any $m$th eigenmode with the corresponding eigenvalue $s_m$ cannot be completely ruled out as unimportant for the synchronizability of the network. This might become particularly important for the case of non-identical laser oscillators where the signless Laplacian matrix $Q$ alone is not determining the results. In fact, in the general case, the spectral gap that is predicted to control the frequency synchronization of non-identical laser models can be defined as the separation between the real parts of the two eigenvalues of matrix $M$ with the smallest real parts, i.e., $\gamma_M = \Re[\lambda_2 - \lambda_1]$. As a result, the spectral criterion  \eqref{k_c} can be approximately extended to non-identical oscillators as 
			\begin{equation}
		\kappa_c =\frac{\bar{b}}{\gamma_M},
		\label{non}
	\end{equation}
	where $\bar{b}$ is a scaling parameter. For the two-oscillator setup of Fig.~1a with frequency detunings $\omega_1 = \omega_0 - \Delta$ and $\omega_2 = \omega_0 + \Delta,$ the spectral gap can be calculated analytically as $\gamma_M = 2 \sqrt{\kappa^2 - \Delta^2},$ yiedling the threshold coupling $\kappa_c=\Delta$ \cite{ding2019dispersive}. Notably, the synchronization threshold is marked with a phase transition in the eigenvalues of matrix $M$ that dictates the system's linearized dynamics (see Supplementary Fig.~1 in the Supplemental Material that also demonstrates this phase transition for random frequency detunings).
	To verify the general approximate criterion \eqref{non} for larger networks, we have numerically calculated the coupling threshold $\kappa_c$ for all possible $21$ network topologies of size $N=5$ and $1,000$ combinations of random frequency detunings. Figure~\ref{fig4} shows that the networks $1,2,$ and $3$, similarly to their identical oscillator counterparts with the zero spectral gap $\gamma=0,$ cannot support the frequency synchronization for any of the chosen frequency detunings. Remarkably, these networks include a locally coupled ring and an all-to-all network representing two opposite ends of the network topology range and are known to be synchronizable in Laplacian oscillator networks \cite{pecora1998master}. It is also worth noting a striking effect that adding one link to the local chain of Fig.~\ref{fig6}top that completes the loop  yields the unsynchronizable ring network~$2$ of Fig.~\ref{fig4}. Out of the remaining $18$ networks with non-zero spectral gap $\gamma,$ and therefore, capable of frequency synchronization according to the spectral criterion, only one, the network $18,$ does not follow the prediction. It remains unsynchronizable for any $\kappa.$ This is the case where a complex interplay between the network structure and distributions of frequency detuning prevents each $m$th eigenmodes with eigenvalue $\lambda_m,$ $m=1,...,N$ from becoming the lasing mode. Nonetheless, as in the identical oscillator case, the spectral criterion singles out the full bipartite network (network 21) as the optimal network topology with the lowest synchronization threshold. 
	To better relate the dependence of the threshold $\kappa_c$ to the identical oscillator criterion \eqref{k_c}, we choose the lowest value of $\kappa_c$ for the full bipartite network (the lowest peak of the corresponding violin plot in Fig.~\ref{fig4}) to identify the lowest scaling constant $\bar{b}$ which could correspond to the least heterogeneous oscillators. We then use this scaling factor via \eqref{k_c}
	for all other networks, see how this trend compares to the actual heterogeneous oscillators (Fig.~\ref{fig4}). Notably, with a few exceptions, even the identical oscillator spectral criterion can predict the general dependence on the spectral network gap $\gamma.$ Obviously, the discrepancy between the predicted trend and the numerical data is due to multiple factors, including non-uniform scaling constants $\bar{b}$ and spectral gaps $\gamma_M$ for different detuning distributions. It is also noteworthy that 
	the spectral gap criterion successfully identifies the full bipartite network as the optimal network topology for frequency synchronization in the Kuramoto-type model obtained from the phase equation in system \eqref{real} by setting $A_n=A_m=1$ (see Supplementary Fig.~2 for the similarities and differences between Fig.~\ref{fig4} and its Kuramoto model counterpart). 
	
	{\it Conclusions.} In this work, we revealed a general approximate principle that relates a critical coupling threshold for  frequency synchronization to the spectral gap between the smallest eigenvalues of the matrix combined from the signless Laplacian connectivity and frequency detuning matrices. The discovered principle demonstrates that the spectral gap of the signless Laplacian, rather than mere connectivity, is a powerful indicator of the synchronizability of such repulsive networks. Although different, this predictive principle may be viewed as an analog of the master stability function for complete synchronization in Laplacian dynamical networks \cite{pecora1998master}, as it isolates, in the identical oscillator case, the contribution of the coupling term from the individual oscillator dynamics. Applying the spectral principle, we discovered that in contrast to one's intuition, both local ring and global network structures prevent frequency synchronization, whereas the fully bipartite network  has optimal synchronization properties. We also demonstrated that this latter property carries over to the repulsive Kuramoto network. The spectral principle has limitations, as it does not always rule out the synchronizability of a complex network of heterogeneous laser oscillators.   
	However, it identifies topologies that can be easily synchronized and used for scalable designs of large laser arrays. Moreover, a maximal spectral gap of the complex matrix incorporating frequency detunings
	could be used as a guiding principle for machine learning approaches to designing disordered laser oscillator networks with optimal synchronization properties required for effective optical computing.
	
	\textit{Acknowledgments.}
	This work was supported by the Air Force Office of Scientific Research (AFOSR) Young Investigator Program (YIP) Award \# FA9550-22-1-0189 (to M.-A.M.) and the Office of Naval Research under Grant No. N00014-22-1-2200 (to I. B.)

%\bibliography{ref.bib}


\end{document}